\documentclass[conference]{IEEEtran}
\IEEEoverridecommandlockouts
\usepackage{cite}
\usepackage{amsmath,amssymb,amsfonts}
\usepackage{algorithmic}
\usepackage{algorithm}
\usepackage{graphicx}
\usepackage{textcomp}
\usepackage{xcolor}
\usepackage{tabularx}
\usepackage{makecell}
\usepackage{caption}
\usepackage{subcaption}
\usepackage{adjustbox}
\usepackage{xcolor}
\usepackage{hyperref}
\DeclareMathOperator*{\argmin}{arg\,min}
\usepackage{soul}
\usepackage{fancyhdr}

\def\BibTeX{{\rm B\kern-.05em{\sc i\kern-.025em b}\kern-.08em
    T\kern-.1667em\lower.7ex\hbox{E}\kern-.125emX}}
\begin{document}

\title{
Digital Twin-Enabled Blockage-Aware Dynamic mmWave Multi-Hop V2X Communication

}

\author{
    \IEEEauthorblockN{Supat Roongpraiwan, Zongdian Li, 
    Tao Yu, and
    Kei Sakaguchi}
    \IEEEauthorblockA{Institute of Science Tokyo, Tokyo 152-8550, Japan.}  
    \IEEEauthorblockA{Emails: \{supat, lizd, yutao, sakaguchi\}@mobile.ee.titech.ac.jp} 

    }

\maketitle

\begin{abstract}
Millimeter wave (mmWave) technology in vehicle-to-everything (V2X) communication offers unprecedented data rates and low latency, but faces significant reliability challenges due to signal blockages and limited range. This paper introduces a novel system for managing dynamic multi-hop mmWave V2X communications in complex blocking environments. We present a system architecture that integrates a mobility digital twin (DT) with the multi-hop routing control plane, providing a comprehensive, real-time view of the network and its surrounding traffic environment. This integration enables the control plane to make informed routing decisions based on rich contextual data about vehicles, infrastructure, and potential signal blockages. Leveraging this DT-enhanced architecture, we propose an advanced routing algorithm that combines high-precision environmental data with trajectory prediction to achieve blockage-aware mmWave multi-hop V2X routing. Our algorithm anticipates network topology changes and adapts topology dynamically to maintain reliable connections. We evaluate our approach through proof-of-concept simulations using a mobility DT of the Nishishinjuku area. Results demonstrate that our DT-enabled routing strategy significantly outperforms conventional methods in maintaining reliable mmWave V2X connections across various traffic scenarios, including fully connected and mixed traffic environments.
\end{abstract}

\section{Introduction}
\label{sec:introduction}
The advent of vehicle-to-everything (V2X) communication has revolutionized intelligent transportation systems (ITS) and connected autonomous vehicles (CAVs). V2X technology establishes a comprehensive network connecting vehicles, pedestrians, and roadside infrastructures through various communication modes: vehicle-to-vehicle (V2V), vehicle-to-pedestrian (V2P), and vehicle-to-infrastructure (V2I). The resulting ecosystem supports a plethora of applications, spanning road safety like forward collision warning (FCW), traffic efficiencies such as green light optimal speed advisory (GLOSA), and commercial services including vehicle near-field payment (VNFP) \cite{cite:v2x-services}. Notably, the safety-centric V2X applications stand out as crucial components in the realization of level 4-5 autonomous vehicles \cite{cite:saej3016}.

In the pursuit of enhanced V2X capabilities, millimeter wave (mmWave) technology has emerged as a significant innovation. Utilizing the expansive bandwidth within the 30 GHz to 300 GHz spectrum, mmWave communication enables multi-gigabit data transfer rates and reduces latency to sub-millisecond levels \cite{cite:mmWave-V2X}. This leap in performance unlocks the potential for advanced V2X applications, with cooperative perception serving as a prime example, facilitating the seamless, real-time exchange of raw sensor data from light detection and ranging (LiDAR) systems and high-resolution cameras \cite{cite:cooperative-perception}.

Despite its promising attributes, integrating mmWave technology into V2X systems presents a significant challenge in ensuring reliable communication \cite{cite:mmwave-v2x-challenge}. Industry standards, as outlined in the 3rd generation partnership project (3GPP) technical specifications TS 22.185 \cite{cite:3gpp.22.185} and TS 22.186 \cite{cite:3gpp.22.186}, require high reliability levels ranging from 90\% to 99.999\% for V2X applications with different complexity. Traditional V2X systems operating at lower frequencies (around 5.9 GHz), such as dedicated short-range communications (DSRC) and long-term evolution (LTE)-V2X, achieve these levels through broadcasting-based routing \cite{cite:v2x-broadcasting}. However, mmWave V2X faces several unique obstacles. Its inherent characteristics, including limited communication range due to high propagation losses, directional transmission, and poor penetration capability, make it particularly susceptible to signal blockages. As a result, conventional broadcasting-based routing strategies are insufficient to maintain high reliability in mmWave V2X networks, necessitating the development of advanced multi-hop management techniques capable of adapting to dynamic traffic conditions.

In light of these challenges, the rapid evolution of internet of things (IoT) technology has introduced a promising innovation called the digital twin (DT). This concept enables real-time monitoring and simulation of physical systems within a virtual environment \cite{cite:DT-trend}. Recent advancements have demonstrated the feasibility of creating mobility DTs that accurately represent dynamic traffic elements such as vehicles, cyclists, and pedestrians, alongside static environmental features like roads, buildings, and trees, all supported by advanced ITS networks \cite{cite:mobility-dt}. Recognizing the potential of these mobility DTs to provide a comprehensive, real-time view of CAV networks with detailed environmental data, this paper proposes a novel approach that employs DT technology to manage mmWave V2X communications in dynamic multi-hop scenarios. Our efforts focused on the following aspects:
\begin{itemize}
    \item We propose an innovative system architecture that utilizes DT technology to enhance multi-hop routing in mmWave V2X networks. Following this, we define the mmWave V2X communication model utilized in our research. Within this framework, we develop an advanced routing algorithm that incorporates trajectory prediction and integrates DT contextualized data, enabling blockage-aware dynamic optimization for V2X routing.
    
    \item To validate the proposed system, we performed a proof-of-concept evaluation using a mobility digital twin of the Nishishinjuku area. Our routing algorithm is assessed in line-of-sight (LOS) V2X scenarios, where we compare its connectivity and communication throughput against conventional routing methods, demonstrating the system’s effectiveness in real-world environments.
\end{itemize}

The remainder of this paper is organized as follows: Section \ref{sec:related-works} presents a critical review of related works, highlighting current limitations in the field. Section \ref{sec:system-design} offers an in-depth exploration of our digital twin-enabled blockage-aware dynamic mmWave multi-hop V2X communication system, including its architecture, communication model, and V2X routing algorithm. Section \ref{sec:system-evaluation} details the evaluation scenario, the system implementation process, and provides a comprehensive evaluation of our proposed system through simulation proof-of-concept. Finally, Section \ref{sec:conclusion} summarizes our findings, conclusions, and outlines promising directions for future research.

\section{Related Works}
\label{sec:related-works}
\subsection{Millimeter-wave routing algorithms for V2X}
Since beamforming is required for mmWave communication to overcome high propagation loss, mmWave communication in vehicular networks faces challenges due to susceptibility to blockages and the need for precise beam alignment. Due to these limitations, mmWave is primarily employed in backhaul networks, which feature a stationary topology. Traditional ad-hoc routing algorithms, such as ad-hoc on-demand distance vector (AODV), dynamic source routing (DSR), and optimized link state routing (OLSR), while functional well in static mmWave mesh networks, often struggle in highly dynamic vehicular environments. Although some researchers have developed routing algorithms that address the specific needs of mmWave-based vehicular networks, including efforts to minimize energy consumption \cite{cite:rw-a-1}, reduce network latency \cite{cite:rw-a-2}, and enable fast link failure recovery \cite{cite:rw-a-3}, these algorithms experience significant performance degradation in mmWave mobile networks, largely due to their lack of consideration for mobility. In \cite{cite:rw-a-4}, a proactive route refinement scheme for mmWave was introduced to alleviate the impacts of mobility and human blockages. However, their simulations were limited to indoor environments, raising concerns regarding their effectiveness in dynamic outdoor settings.

Notably, some recent works on routing algorithms for mmWave-based vehicular networks have considered beam alignment issues \cite{cite:rw-a-5, cite:rw-a-6}. Nonetheless, these approaches typically rely on accurate vehicle position information, which can be challenging to obtain reliably in real time due to CSMA/CA-induced delays in traditional vehicular ad hoc networks (VANETs) protocol. The limitations of existing approaches highlight the need for more sophisticated, context-aware routing solutions that can handle practical V2X environments, particularly in terms of blockage awareness and dynamic adaptation.

\subsection{Software-defined vehicular networks}
The decentralized network architecture of traditional VANETs presents significant challenges in collecting global information, such as node location and status, which is crucial for multi-hop routing and topology acquisition. To address these limitations, software-defined networking (SDN) has emerged as a promising solution. Its centralized architecture significantly enhances network control plane functions, such as routing management, by providing greater visibility, robustness, and responsiveness to the data plane dynamics \cite{cite:rw-b-1}. Consequently, several researchers have proposed software-defined vehicular networks (SDVNs) as a novel V2X network architecture \cite{cite:rw-b-2, cite:rw-b-3, cite:rw-b-4}. The development of SDVNs has progressed beyond theoretical and simulation levels, with researchers building SDVN testbeds \cite{cite:rw-b-5, cite:rw-b-6} and conducting field trials for SDVN-based multi-hop mmWave V2X \cite{cite:rw-b-7, cite:rw-b-8}.

While SDVNs provide an efficient means to generate and update global network topology, they are limited to information from connected entities such as CAVs and roadside units (RSUs). This limitation is particularly problematic for mmWave multi-hop routing, where non-connected vehicles and other physical objects can significantly impact V2X link quality and availability. To address this gap, the aforementioned concept of DT emerges as a crucial component in mmWave V2X systems. DT incorporates detailed 3D models of all relevant elements, including both connected and non-connected entities, as well as dynamic and static objects. Thus, DT technology is essential to further optimize the performance and reliability of mmWave V2X systems beyond the limitations of SDVNs.

\section{System Design}
\label{sec:system-design}
\subsection{System architecture}
\begin{figure}[tb] 
    \includegraphics[width=\linewidth]{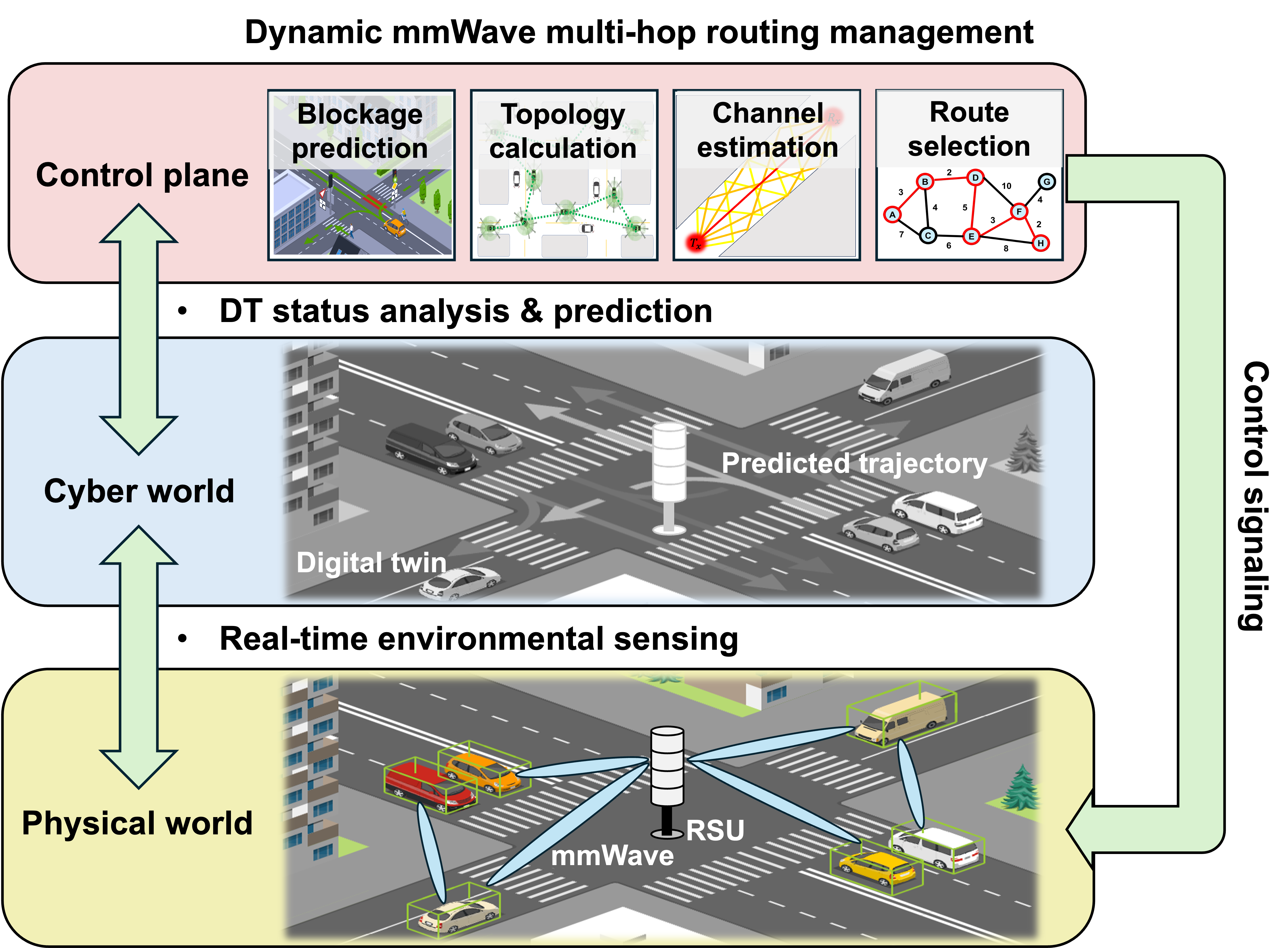} 
    \centering 
    \caption{System architecture.} 
    \label{fig:sys-arch} 
\end{figure}

This section introduces an advanced system architecture that leverages DT technology to enable blockage-aware, dynamic multi-hop mmWave V2X communication in complex urban environments. As illustrated in Fig. \ref{fig:sys-arch}, this system architecture provides an intelligent V2X control plane for managing mmWave multi-hop routing, utilizing real-time dynamic information and blockage prediction from the DT.

The DT of the vehicular system consists of both the physical world and the cyber world. The cyber world model is initially constructed using location-specific static data, including high-definition 3D maps, models of buildings and the environment, and vector maps containing detailed traffic and road infrastructure information. As real-time sensor data are gathered from RSUs and CAVs, they are transmitted to the DT server via V2X communication. To ensure accurate synchronization, time synchronization is applied at all sensing and communication nodes as well as the DT servers. This dual-sided synchronization aligns sensor data with the DT server’s reference time, minimizing the potential impact of transmission delays and ensuring that real-time vehicle status, including their positions and dimensions, is accurately reflected in the DT.

In this study, basic location-specific data, such as GPS coordinates, is used to update the DT with real-time vehicle data. Meanwhile, 3D maps and building models, which are commonly available from open-source maps, are used for LOS detection in the proposed routing algorithm. In cases where static location-specific data is unavailable, alternative solutions can be employed. Real-time sensor fusion from onboard vehicle sensors and cooperative perception from connected vehicles can help compensate for missing static data and facilitate the construction of a partial DT \cite{cite:mobility-dt, cite:cooperative-perception, cite:coop-perception-for-dt}. The integration of static and dynamic data ensures that the DT cyber world can accurately represent the behavior of vehicles in the physical world, as well as their surroundings and potential communication obstacles in real-time.

The blockage-aware V2X control plane leverages comprehensive vehicle and environmental information in the DT cyber world to simulate the mmWave channel model for connectivity assessment, anticipate potential blockages using predictive machine learning, and select the most efficient communication routes for current and future data transmission. The mmWave channel model used in this research is detailed in Section \ref{sec:channel-model}. By employing advanced machine learning techniques based on historical data and environmental factors, the control plane can proactively adjust V2X communication routing to ensure consistent connectivity, even in highly dynamic scenarios. A thorough explanation of the proactive blockage-aware multi-hop routing algorithm used in the control plane, which considers DT latency and mitigates its impact on routing decisions, can be found in Section \ref{sec:routing-algo}.

By integrating real-time information exchange within the DT and the automated blockage-aware V2X control plane, the proposed system can ensure reliable, dynamic, and blockage-resilient routing for the mmWave V2X network. This approach significantly enhances the reliability and efficiency of mmWave communication in challenging dynamic environments where LOS obstructions are common and frequently changing.

\subsection{Communication model}
\label{sec:channel-model}

This section outlines the mmWave V2X communication model employed in this research to assess the channel status of mmWave V2X links in dynamic traffic environments. Our channel model builds upon the 3GPP technical report TR 37.885 \cite{cite:3gpp.37.885}, focusing specifically on LOS V2X scenarios. The path loss model for V2X LOS link is given by:
\begin{equation}
   \text{PL}(d)[dB] = \alpha + \beta \log_{10}(f_{\text{mmWave}}) + \gamma \log_{10}(d) + X_\text{SF}.
\label{eq:pathloss}
\end{equation}
The model parameters are defined as follows: $\alpha$ represents the constant path loss factor, $\beta$ characterizes the frequency-dependent path loss exponent, $f_{\text{mmWave}}$ denotes the millimeter wave center frequency in GHz, $\gamma$ represents the distance-dependent path loss exponent, and $d$ is the V2X link distance in meters. The term $X_\text{SF}$ accounts for shadow fading effects from building structures, which follows a normal distribution $\mathcal{N}(0,\sigma_\text{SF})$. In this path loss model, the first two terms characterize free-space path loss, whereas the third term accounts for distance-based attenuation beyond 1 meter.

To effectively employ blockage avoidance routing for dynamic obstacles, such as in non-line-of-sight caused by vehicle (NLOSv) scenarios, we incorporate a blocking loss model that follows a normal distribution:
\begin{equation}
\begin{split}
   \text{BL}(d)[\text{dB}] &= \mathcal{N}(\mu_{offset} + \mu_d(d), \sigma_\text{BL}), \\
   \text{where} \quad \mu_d(d) &= \max(0, \xi \log_{10}(d) - \text{BL}_{thresh}).
\end{split}
\label{eq:blockingloss}
\end{equation}
In this model, $\mu_{offset}$ represents the blocking loss mean offset, $\xi$ characterizes the blocking loss exponent, $d$ is the V2X link distance in meters, $\text{BL}_{thresh}$ is a blocking loss threshold parameter in the $\mu_d$ calculation where the max function ensures non-negative values, and $\sigma_{BL}$ denotes the standard deviation of the blocking loss.

\begin{table}[t]
\centering
\caption{Parameters value of the mmWave V2X communication model.}
\label{tab:channel_parameters}
\begin{tabularx}{1.0\linewidth}{|X|>{\centering\arraybackslash}p{1.5cm}|>{\centering\arraybackslash}p{1.5cm}|}
\hline
\textbf{Parameters} & \textbf{Symbol} & \textbf{Value} \\
\hline
Path loss constant & $\alpha$ & 38.77 dB \\
\hline
Frequency-dependent path loss exponent & $\beta$ & 16.7 \\
\hline
Distance-dependent path loss exponent & $\gamma$ & 18.2 \\
\hline
Shadow fading standard deviation & $\sigma_\text{SF}$ & 3 dB \\
\hline
Blocking loss mean offset & $\mu_{offset}$ & 9 dB \\
\hline
Blocking loss exponent & $\xi$ & 15 \\
\hline
Blocking loss threshold & $\text{BL}_{thresh}$ & 41 dB \\
\hline
Blocking loss standard deviation & $\sigma_{\text{BL}}$ & 4.5 dB \\
\hline
\end{tabularx}
\end{table}

The parameter values for both path loss and blocking loss models are provided in Table \ref{tab:channel_parameters}. These values were specifically calibrated for urban V2X scenarios based on the environmental characteristics outlined in Section \ref{sec:scenario}. This channel model is universal across all frequencies within the mmWave band and serves as the foundation for our proposed DT-based blockage-aware V2X routing system.

\subsection{DT-based blockage-aware routing algorithm}
\label{sec:routing-algo}

\begin{figure*}[tb] 
    \includegraphics[width=0.8\linewidth]{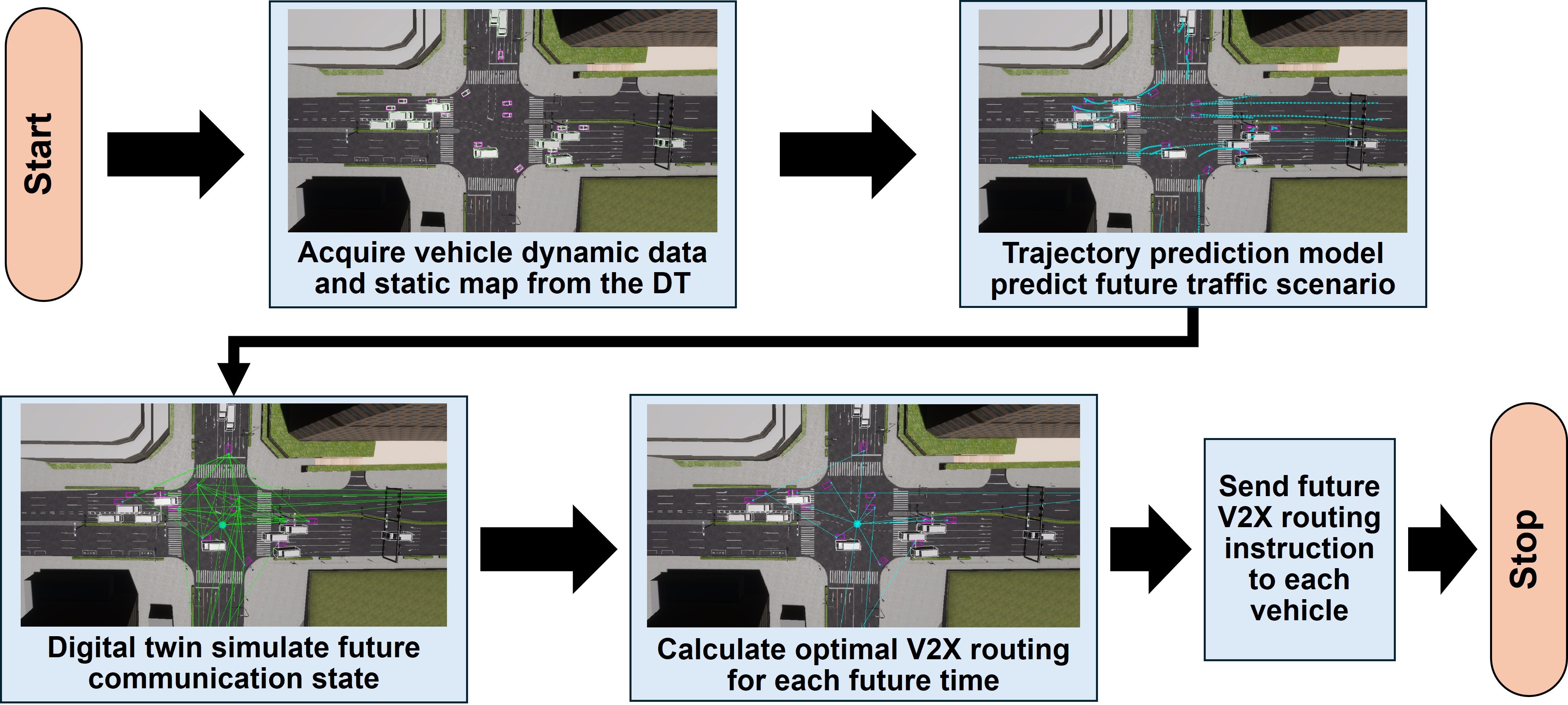} 
    \centering 
    \caption{DT-based blockage-aware routing algorithm flowchart.} 
    \label{fig:routing-flowchart} 
\end{figure*}

\begin{table}[t]
\centering
\caption{Time scale parameters value of the DT-based blockage-aware routing algorithm.}
\label{tab:time_scale_parameters}
\begin{tabularx}{1.0\linewidth}{|X|>{\centering\arraybackslash}p{2cm}|}
\hline
\textbf{Parameters} &  \textbf{Value} \\
\hline
Base timestep duration &  100 ms \\
\hline
V2X topology control interval &  1 timestep \\
\hline
Routing algorithm execution interval &  10 timesteps \\
\hline
Historical data window for trajectory prediction &  10 timesteps \\
\hline
Future topology prediction horizon &  50 timesteps \\
\hline

\end{tabularx}
\end{table}

This section presents a DT-based blockage-aware routing algorithm designed for the control plane of our system architecture. The routing algorithm aims to determine optimal V2X communication paths while considering both signal propagation and potential blockages by surrounding vehicles.

The flowchart in Fig. \ref{fig:routing-flowchart} outlines the flow of the proposed routing algorithm, which operates in constant time. The process begins by collecting real-time vehicle data and static map information from the digital twin system. This information is then fed into a trajectory prediction model, which forecasts future traffic scenarios. Based on these predictions, the digital twin simulates potential communication states. Using this simulated data, the system determines the optimal V2X routing for each future time point using the V2X topology calculation algorithm (Algorithm \ref{algo:topology}). Finally, these routing instructions are transmitted to every vehicle in the network, enabling proactive optimization of V2X communications and anticipating potential issues before they occur. By running the process in constant time, the system can efficiently plan future V2X topology, providing robustness to DT-vehicle latency, a factor that exists in real-world environments. This proactive approach optimizes V2X communications and anticipates potential issues before they occur.

The time scale characteristics of the routing algorithm implementation are presented in Table \ref{tab:time_scale_parameters}. The system operates with a base timestep of 100 ms for dynamic topology control. The V2X topology is updated in each timestep. The proposed routing algorithm executes at every 10 timesteps interval (1 s) and leverages the trajectory prediction model, which processes 10 timesteps of historical vehicle data to forecast future traffic states and calculate optimal V2X topology for the next 50 timesteps. This predictive approach enables the algorithm to pre-calculate V2X topology configurations, effectively decoupling the topology implementation from the routing algorithm's computation time. The pre-calculated topology configurations can be deployed instantly as needed, ensuring seamless V2X communication optimization.

\begin{algorithm}[t]
\caption{V2X topology calculation algorithm}
\begin{algorithmic}[1]
\FORALL{LOS connection $(s,t)$ in connection graph}
    \STATE Calculate $pathLoss$ from Eq. \eqref{eq:pathloss}
    \STATE Compute $linkVector$ from source $S$ to destination $D$ antenna 
    \STATE Instantiate $memory$ for storing potential $BRF$
    \FORALL{vehicle $V$ in scene}
        \IF{$V \neq S$ and $V \neq D$}
            \STATE Calculate $blockingPerpendicularDistance$ from vehicle $V$ to $linkVector$
            \IF{$V$ potentially blocks $linkVector$}
                \STATE Get all prediction error $\varepsilon$ from prediction error heatmap
                \STATE Calculate potential $BRF$ from Eq. \eqref{eq:blocking-prob}
                \STATE Store potential $BRF$ in the $memory$
            \ENDIF
        \ENDIF
    \ENDFOR
    \STATE Get maximum $BRF$ from $memory$
    \STATE Calculate $blockingLoss$ from Eq. \eqref{eq:blockingloss}
    \STATE Calculate $connectionWeight$ from Eq. \eqref{eq:weight}
\ENDFOR

\FORALL{$(S,D)$ in V2X demands}
    \STATE Calculate optimal topology $\tau_n^*(S \to D)$ from Eq. \eqref{eq:topology}
    \STATE Add $\tau_n^*(S \to D)$ to $v2xTopology$
\ENDFOR

\RETURN $v2xTopology$
\end{algorithmic}
\label{algo:topology}
\end{algorithm}

The V2X topology calculation using Algorithm \ref{algo:topology} is divided into two distinct parts: the calculation of communication link weight for each available LOS link (line 1-18) and the selection of the optimal communication topology for V2X communications (line 18-23). This process is repeated to determine the optimal V2X topology for each future timestep based on the trajectory prediction results.

\paragraph{Communication link weight calculation}

The communication link weight for each V2X link at each future timestep is determined based on the predicted future traffic state from the trajectory prediction model. Relying solely on mmWave signal path loss (PL) can make the system overly dependent on the accuracy of the prediction model, which may not always be reliable. To mitigate this, the research introduces an additional value called the blocking risk factor (BRF), which incorporates the model's prediction error into the assessment of blocking risks. By combining BRF with blocking loss (BL), the system can assess the potential blocking risks and capture a more comprehensive view of the overall communication link quality, enhancing its resilience to the inaccuracies in the trajectory prediction model. This approach ensures more robust and adaptive communication link weight calculations for future V2X communication scenarios. The communication link weight can be calculated by the sum of PL and the potential blocking loss (product of $\lambda$, BRF, and BL), as shown in Eq. \eqref{eq:weight}:
\begin{equation} W_{connection} = \text{PL} + \lambda \cdot  \text{BRF} \cdot \text{BL},
\label{eq:weight}
\end{equation}
where $\lambda$ denotes a communication link weight hyperparameter that controls the contribution of blocking-related factors (BRF and BL) to the overall communication link weight.

\begin{figure}[tb] 
    \includegraphics[width=\linewidth]{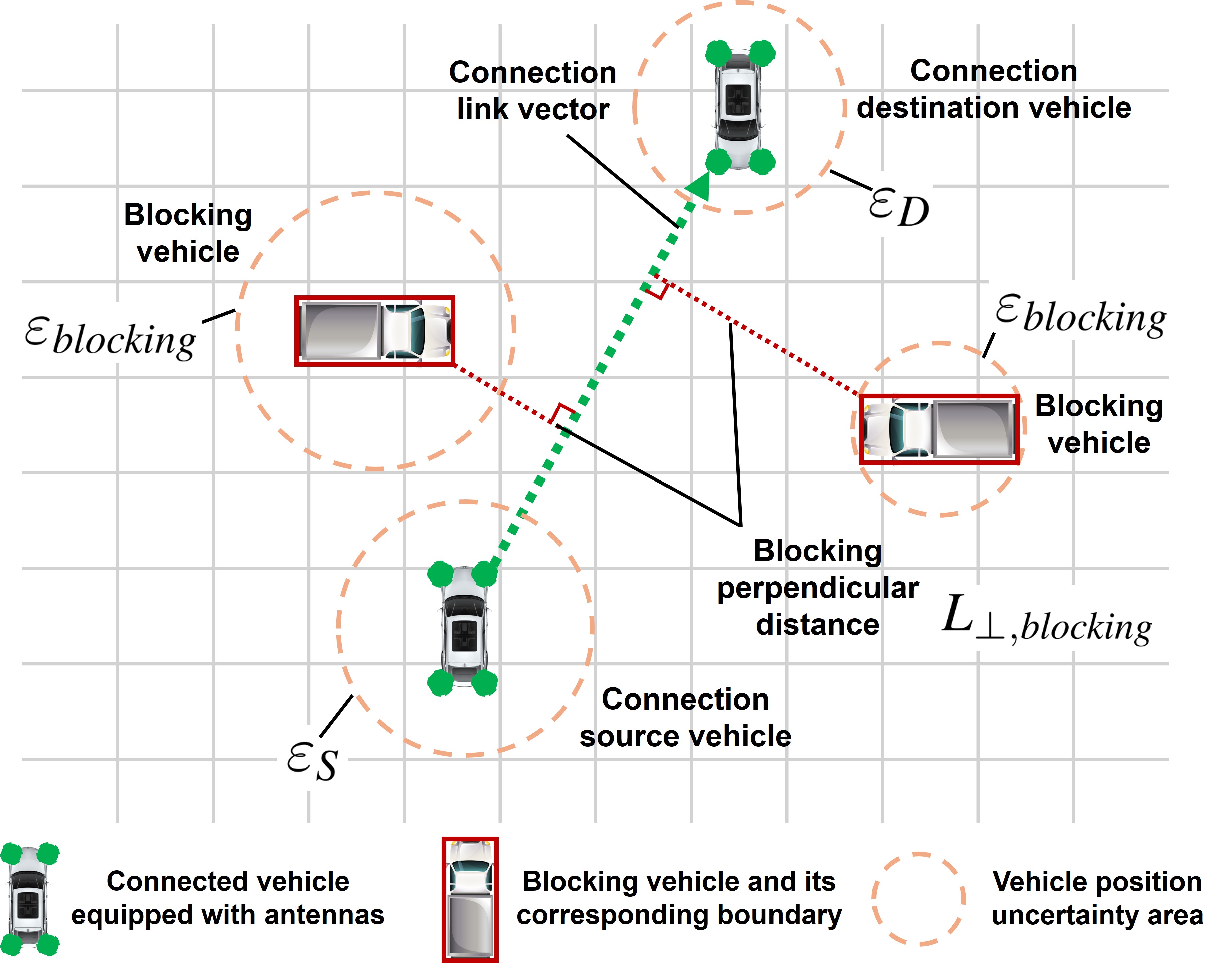} 
    \centering 
    \caption{Blocking risk factor calculation.} 
    \label{fig:blocking-prob} 
\end{figure}

From Algorithm \ref{algo:topology}, in order to calculate the weight connection, the PL and BL can be derived from the distance of V2X link using Eq. \eqref{eq:pathloss} and Eq. \eqref{eq:blockingloss}. On the other hand, BRF is defined as the maximum ratio of combined prediction errors to the perpendicular distance of potential obstacles, expressed as:
\begin{equation} \text{BRF} = \max \left( \frac{\varepsilon_S + \varepsilon_D + \varepsilon_{blocking}}{L_{\perp, blocking}} \right)
\label{eq:blocking-prob}
\end{equation}
In this equation, $\varepsilon_S$,  $\varepsilon_D$, and $\varepsilon_{blocking}$ represent the model prediction errors for the connection source vehicle, connection destination vehicle, and potential blocking vehicles, respectively, while $L_{\perp, blocking}$ denotes the perpendicular distance of blocking vehicle to the connection link vector as illustrated in Fig. \ref{fig:blocking-prob}. 

\begin{figure}[t]
    \centering
    \begin{subfigure}{0.4\textwidth}
        \includegraphics[width=\linewidth]{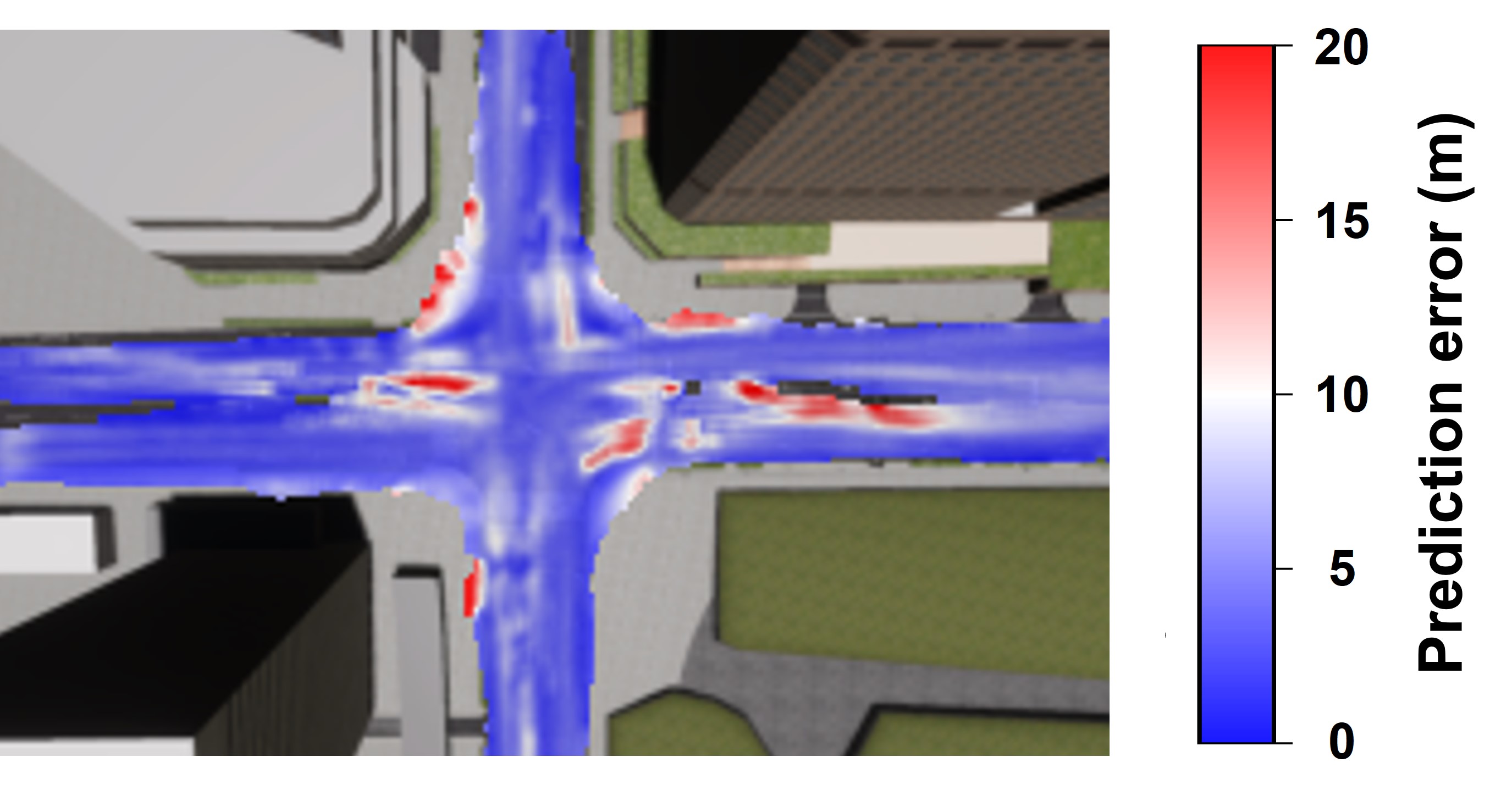}
        \caption{Prediction error heatmap of initial model}
        \label{fig:heatmap-before}
    \end{subfigure}
    \hfill
    \begin{subfigure}{0.4\textwidth}
        \includegraphics[width=\linewidth]{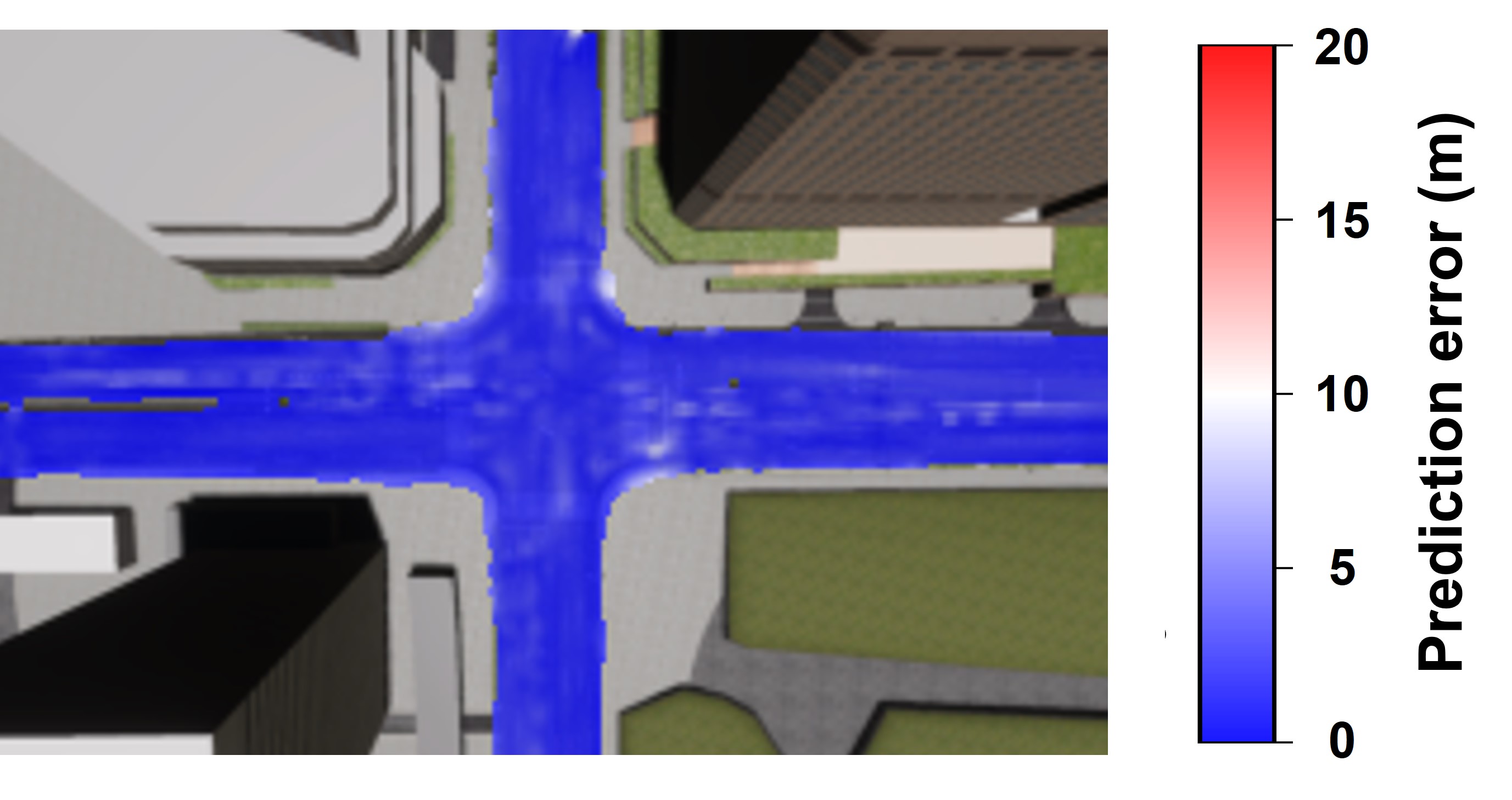}
        \caption{Prediction error heatmap of model after continuous learning by digital twin}
        \label{fig:heatmap-after}
    \end{subfigure}
    \caption{Dynamic heatmap of trajectory prediction model's prediction error.}
    \label{fig:dynamic-map}
\end{figure}

The prediction error can be obtained from the DT's dynamic heatmap of the trajectory prediction model's prediction error (Fig. \ref{fig:dynamic-map}). This heatmap represents the average prediction error for each vehicle position and is continuously updated using DT's real-time data on prediction results and actual vehicle positions. To continuously enhance the model's accuracy, we implement a continuous learning mechanism based on selective fine-tuning. Specifically, when high-error regions are detected in the heatmap, the model prioritizes these areas for targeted retraining by increasing the ratio of data from these weak spots in the fine-tuning process. This adaptive learning approach enables the model to incrementally refine its predictions, as illustrated in Fig. \ref{fig:heatmap-after}, which demonstrates a reduction in prediction errors and improved consistency across all regions. By integrating this DT-enabled dynamic heatmap, our system not only provides real-time prediction error data for communication link weight calculation but also ensures ongoing optimization of the trajectory prediction model through targeted continuous learning.

By applying this method to all available LOS links, we can calculate the communication link weight for each potential V2X connection. This process results in a new weighted connection graph, where each edge represents a possible communication link with its associated communication link weight. This weighted graph serves as the foundation for the subsequent optimal V2X topology selection.
\paragraph{Optimal V2X topology selection}
After calculating the communication link weights for all LOS links, the next step is to determine the optimal V2X communication topology. This is achieved using the following equation:
\begin{equation} \tau_n^*(S \to D) = \argmin_{\tau \in T_n(S \to D)} \sum_{u,v \in \tau} W_{\text{connection}}(u,v)
\label{eq:topology}
\end{equation}
Eq. \eqref{eq:topology} defines the optimal topology $\tau_n^*(S \to D)$ for a communication path from source ($S$) to destination ($D$). The equation seeks to minimize the sum of communication link weights along the path, considering all possible topologies $T_n(S \to D)$ with $n$ multi-routes. In a multi-hop scenario, each route may consist of one or more relay ($R$) vehicles between $S$ and $D$, forming paths like $S \to R \to D$ or $S \to R_1 \to R_2 \to D$.

To solve this equation, we can employ different algorithms depending on the value of $n$. For single-route scenarios ($n = 1$), the Dijkstra's shortest path algorithm \cite{cite:dijkstra} can be used to find the optimal route. To further enhance reliability, this research also employs repetition-based multi-route diversity \cite{cite:multiroute} for $n > 1$ scenarios, where the Yen's algorithm \cite{cite:yen-algo} is used to determine the top-$n$ shortest paths. This multi-route diversity enhances system connectivity since the connection remains active as long as at least one route maintains connectivity. The system repeats this calculation for each future timestep and sends routing instructions to all connected vehicles.

In the scene of complexity, we can analyze the complexity of the algorithm in each part. Firstly, the communication link weight calculation part iterates through all LOS connections in the connection graph, where for each connection, it examines all vehicles as potential blockers. With N vehicles, assuming the number of connections is $O(N)$ in a realistic sparse V2X network, and each connection needs to check against $O(N)$ potential blocking vehicles, this part has a complexity of $O(N^2)$. Next, the optimal V2X topology selection part processes V2X demands using Yen's K-shortest path algorithm, which has a complexity of $O(K\cdot N \cdot(N+N\log N))=O(KN^2\log N)$ \cite{cite:yen-algo}, where K is the number of multi-route. Therefore, the overall complexity of the proposed V2X topology calculation algorithm is $O(N^2) + O(KN^2\log N) = O(KN^2\log N)$.  Since Yen's algorithm dominates the complexity, this shows that the proposed algorithm maintains computational efficiency and remains practical for real-world V2X deployments.

In summary, the DT-based blockage-aware routing algorithm effectively enhances V2X communication by integrating real-time vehicle data and static map information into a dynamic framework. By leveraging trajectory prediction models and incorporating BRF into communication link weight calculations, the algorithm provides a robust approach to optimize communication paths while proactively addressing potential blockages, allowing for efficient and adaptive routing solutions that are resilient to the inherent uncertainties in real-world vehicular environments. This proactive strategy ultimately contributes to the overall reliability and performance of V2X communications, ensuring seamless connectivity in evolving traffic scenarios.

\section{System Evaluation}
\label{sec:system-evaluation}
\subsection{Scenario}
\label{sec:scenario}

To evaluate the efficacy of the proposed system architecture, we have designed an experimental scenario focusing on LOS mmWave communication in urban environments. By emphasizing LOS communication, the research aims to evaluate the system’s ability to maintain reliable V2X connections in the presence of physical obstacles and dynamic conditions.

\begin{figure}[tb] 
    \includegraphics[width=\linewidth]{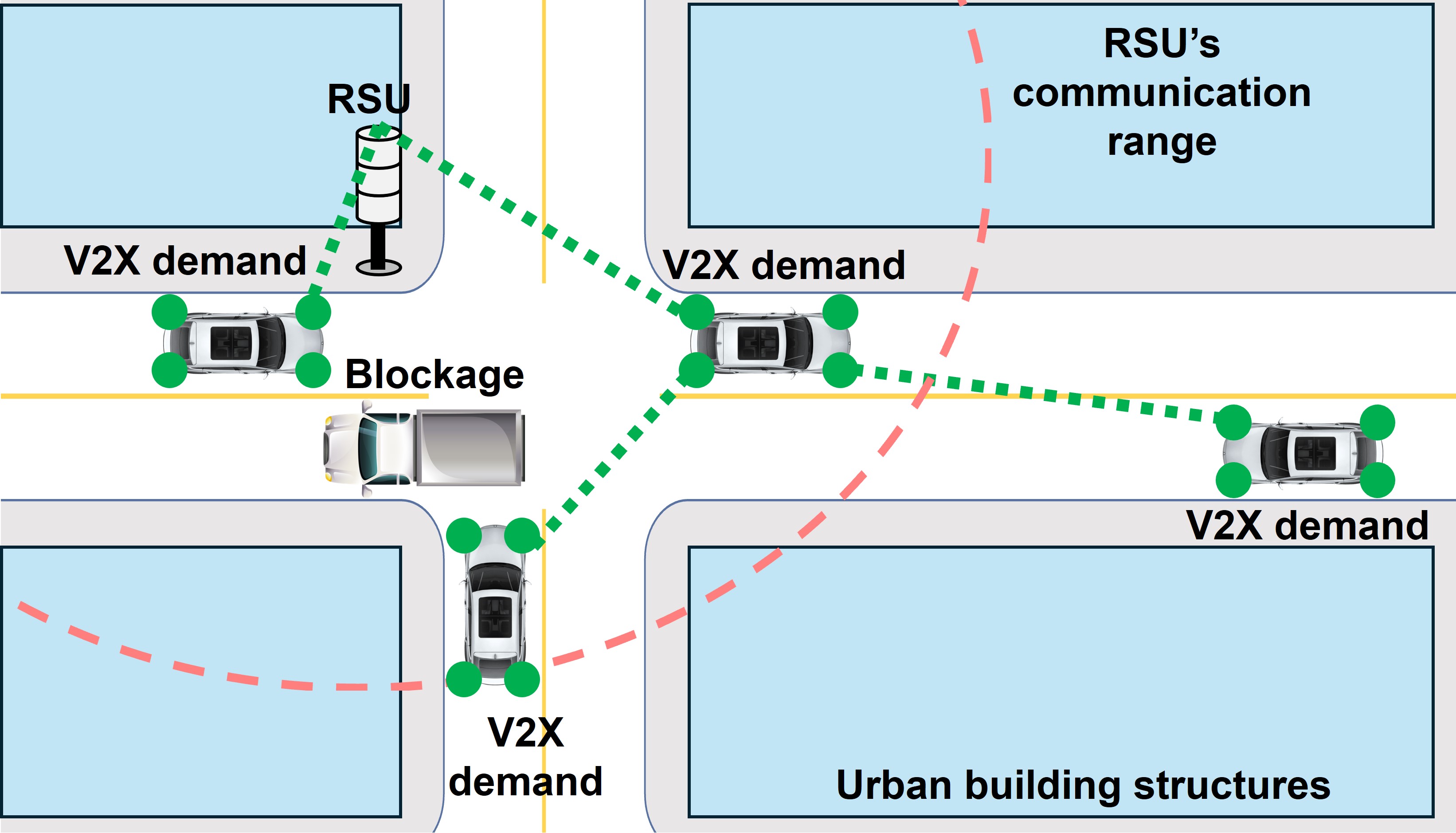} 
    \centering 
    \caption{System evaluation scenario.} 
    \label{fig:scenario} 
\end{figure}

Fig. \ref{fig:scenario} illustrates our system evaluation scenario, which models a single urban intersection. An RSU is positioned at a height of 5 meters in the center of the intersection, serving as the central node for V2X communication \cite{cite:rsu-height}. The scenario incorporates both connected and unconnected vehicles navigating the intersection. Each connected vehicle is equipped with four multiple input multiple output (MIMO) antennas, positioned at each top corners of the vehicle, enabling multiple connections for each antenna to enhance communication reliability. The goal for connected vehicles is to establish V2X communication with the RSU, either through a direct single-hop connection or by employing multi-hop communication via other connected vehicles when obstacles block LOS or vehicles are beyond the RSU's single-hop connection range.

To evaluate the system's performance under different conditions, proof-of-concept experiments were conducted in two key traffic scenarios:
\paragraph{Fully connected traffic scenario} This scenario assumes that all vehicles in the simulation are fully equipped with V2X communication capabilities, enabling an evaluation of system performance under ideal conditions.
\paragraph{Mixed traffic scenario} In this scenario, a mix of connected and unconnected vehicles is introduced, reflecting real-world conditions where not every vehicle is equipped with V2X communication technology. Unconnected vehicles act as mobile obstacles that can block LOS communication paths, requiring the system to dynamically adapt by establishing multi-hop communication links.

These distinct scenarios provide insights into the system's performance under both optimal and challenging conditions. By comparing results from these two scenarios, we can assess the system's robustness, adaptability, and effectiveness in maintaining V2X connectivity in diverse urban traffic environments.

\subsection{Evaluation Metric}
To assess the system’s performance, we define the connectivity metric as follows:

\begin{equation} Connectivity = \frac{\sum_{t=1}^{T} CV_{\text{successful}, t}}{\sum_{t=1}^{T} CV_{\text{total}, t}} \label{eq:connectivity
} \end{equation}
In this expression, $CV_{\text{successful}, t}$ denotes the number of connected vehicles that successfully achieve their V2X communication demands at each timestep $t$, whether through direct communication links or multi-hop communication. While, $CV_{\text{total}, t}$ represents the total number of connected vehicles within the V2X network at timestep $t$, regardless of their connectivity status. This metric provides a comprehensive evaluation of system connectivity by quantifying the ratio of successfully connected vehicles to the total number of vehicles in the network over the entire observation period, offering a measure of the system's effectiveness in ensuring reliable V2X communication.

\subsection{Simulation Implementation}
\label{sec:sim-implement}

\begin{figure}[tb]
    \includegraphics[width=\linewidth]{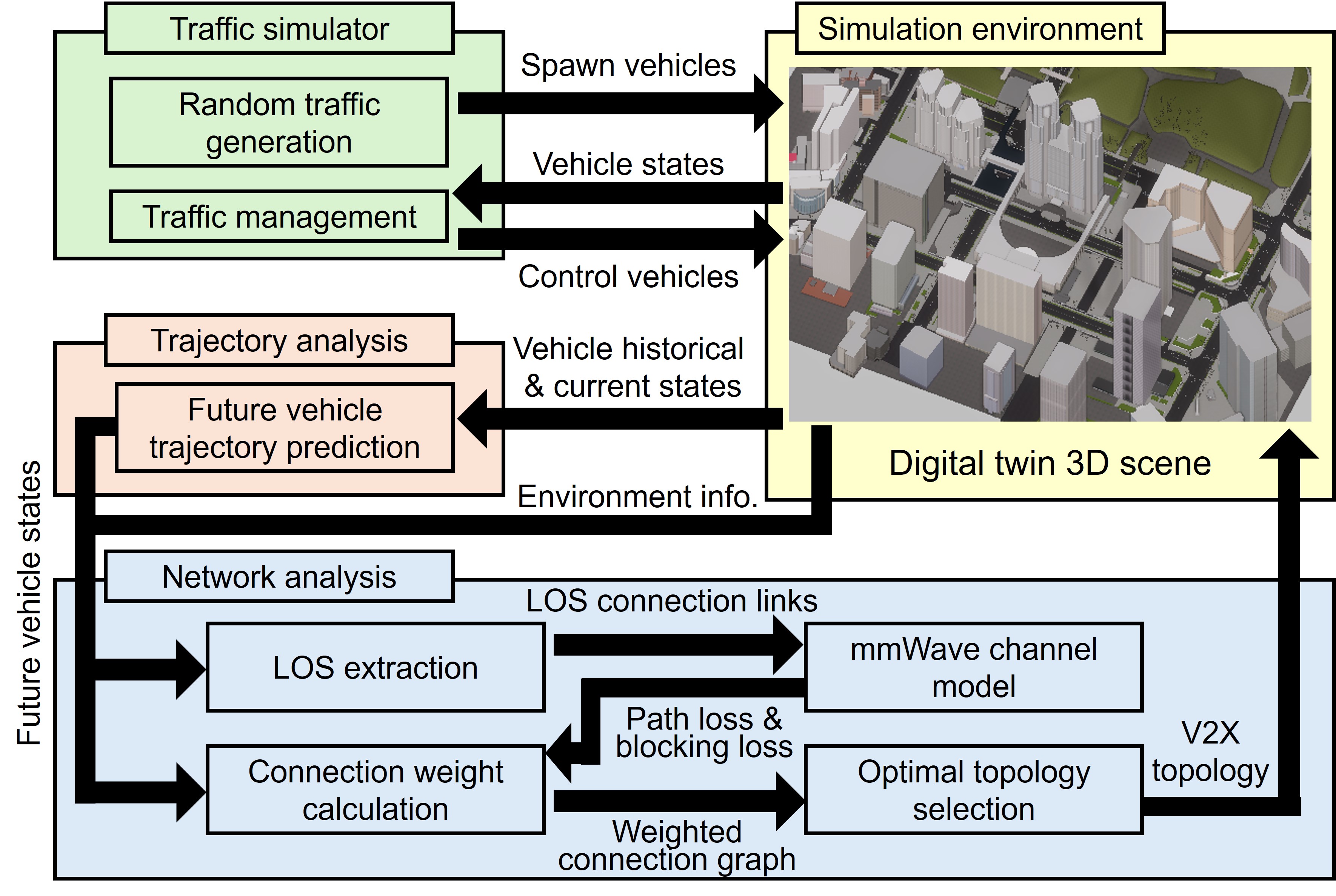}
    \centering
    \caption{Functional diagram of simulation system implementation.}
    \label{fig:sys-implement}
\end{figure}

The proposed system is implemented through an integrated framework as illustrated in Fig. \ref{fig:sys-implement}, which consists of a traffic simulator, trajectory analysis module, and network analysis module, along with a digital twin 3D scene representing the simulation environment.

The simulation environment is developed in Unity and replicates the urban landscape of the Nishi-shinjuku area. For realistic traffic simulation, we incorporate the AWSIM package \cite{cite:awsim}, an open-source autonomous driving simulator that manages vehicle spawning, state transitions, and control within the simulation environment.

For trajectory analysis, we implement a Python module that processes historical and real-time vehicle states from the simulation. At its core is a sophisticated multi-layer long short-term memory (LSTM) model \cite{cite:lstm} that predicts future vehicle trajectories. The LSTM architecture comprises five layers with 64 neurons each, optimizing the balance between computational efficiency and prediction accuracy.

The network analysis component leverages Unity's ray-casting to detect LOS connections between antennas. These LOS links serve as inputs to the mmWave channel model detailed in Section \ref{sec:channel-model}. The system constructs weighted connection graphs using Eq. \eqref{eq:weight}, incorporating path loss and blocking loss from the mmWave channel model along with the BRF derived from trajectory prediction results. The optimal V2X network topology is then determined using shortest path algorithms from the NetworkX package \cite{cite:networkx}.

The entire system is integrated using the robot operating system 2 (ROS2) \cite{cite:ros2}, which facilitates communication between Unity and Python components.

For comparison, we also implement two conventional routing algorithms: an SDVN approach that selects the best communication route based on the positions of vehicle and RSU antennas, and a single-hop method that only selects direct communication to the RSU based on available communication links. The SDVN is a state-of-the-art benchmark due to its efficient vehicular communication via centralized management with a control plane \cite{cite:rw-b-8, cite:sdvn-sota, cite:sdvn-geo-pos}. The single-hop method serves as a baseline in mmWave V2X environments where direct transmission is preferred for its high data rates and low latency in LOS conditions \cite{cite:single-hop-sota}.

\begin{figure}[tb]
    \includegraphics[width=\linewidth]{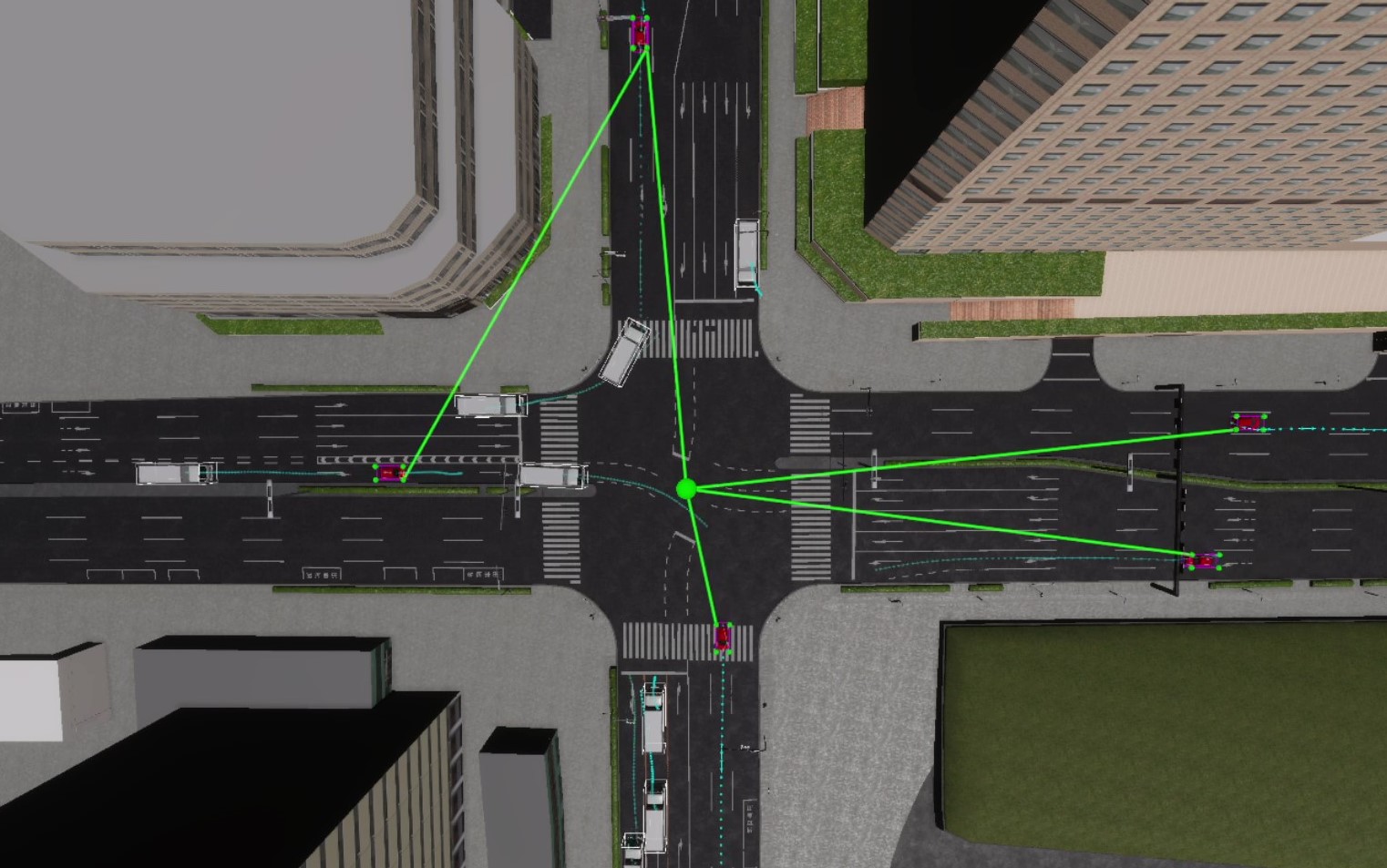}
    \centering
    \caption{Snapshot of the simulation. Connected vehicles are shown in red and blocking vehicles are shown in white. The blue lines represent trajectory prediction results. The green sphere indicates antenna positions of vehicles and RSU. Green lines display the V2X communication topology.}
    \label{fig:sim-snapshot}
\end{figure}

Fig. \ref{fig:sim-snapshot} shows a snapshot of the simulation in action, demonstrating the system's capabilities. In the visualization, connected vehicles are in red while blocking vehicles are shown in white. The blue lines represent trajectory prediction results from the LSTM model. The green sphere indicates antenna positions of vehicles and RSU, while green lines display the optimized V2X communication topology determined by the network analysis module.

\subsection{Results}
\begin{figure}[t]
    \centering
    \includegraphics[width=\linewidth]{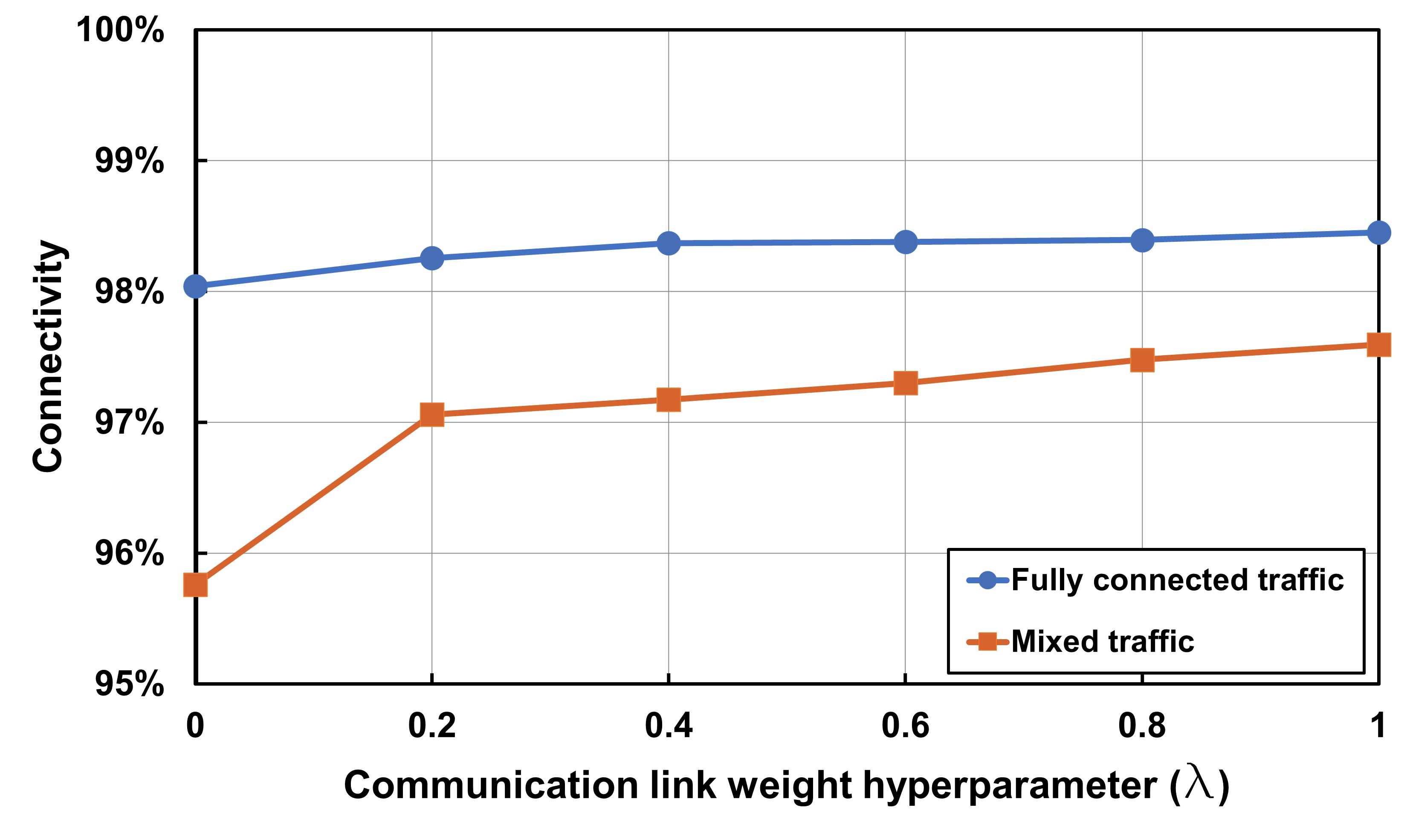}
    \caption{Impact of weight hyperparameter on connectivity.}
    \label{fig:connectivity-lambda}
\end{figure}

\begin{figure}[t]
    \centering
    \includegraphics[width=\linewidth]{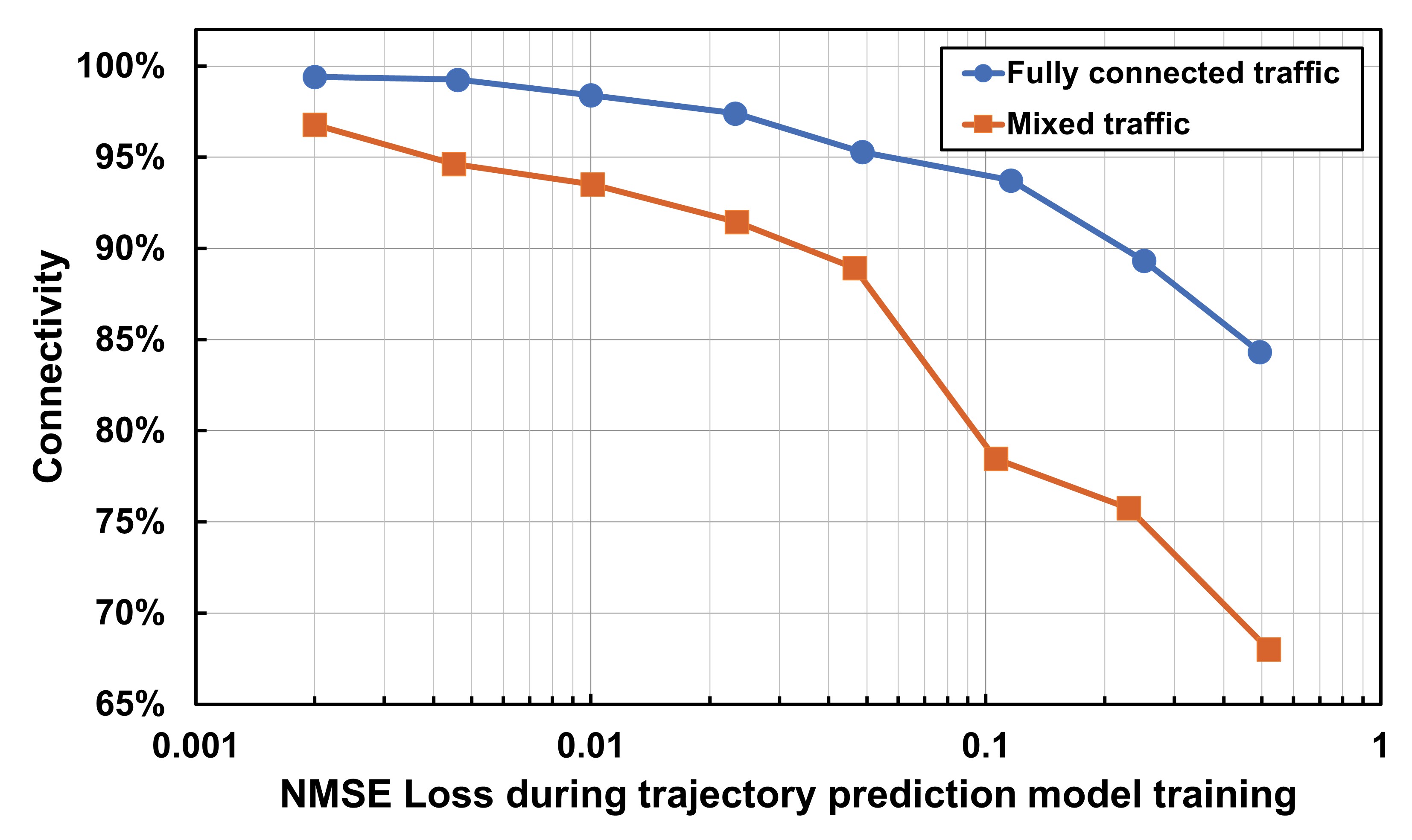}
    \caption{Relationship between trajectory prediction model's NMSE training loss and system connectivity.}
    \label{fig:model-loss}
\end{figure}

\begin{figure}[t]
    \centering
    \includegraphics[width=\linewidth]{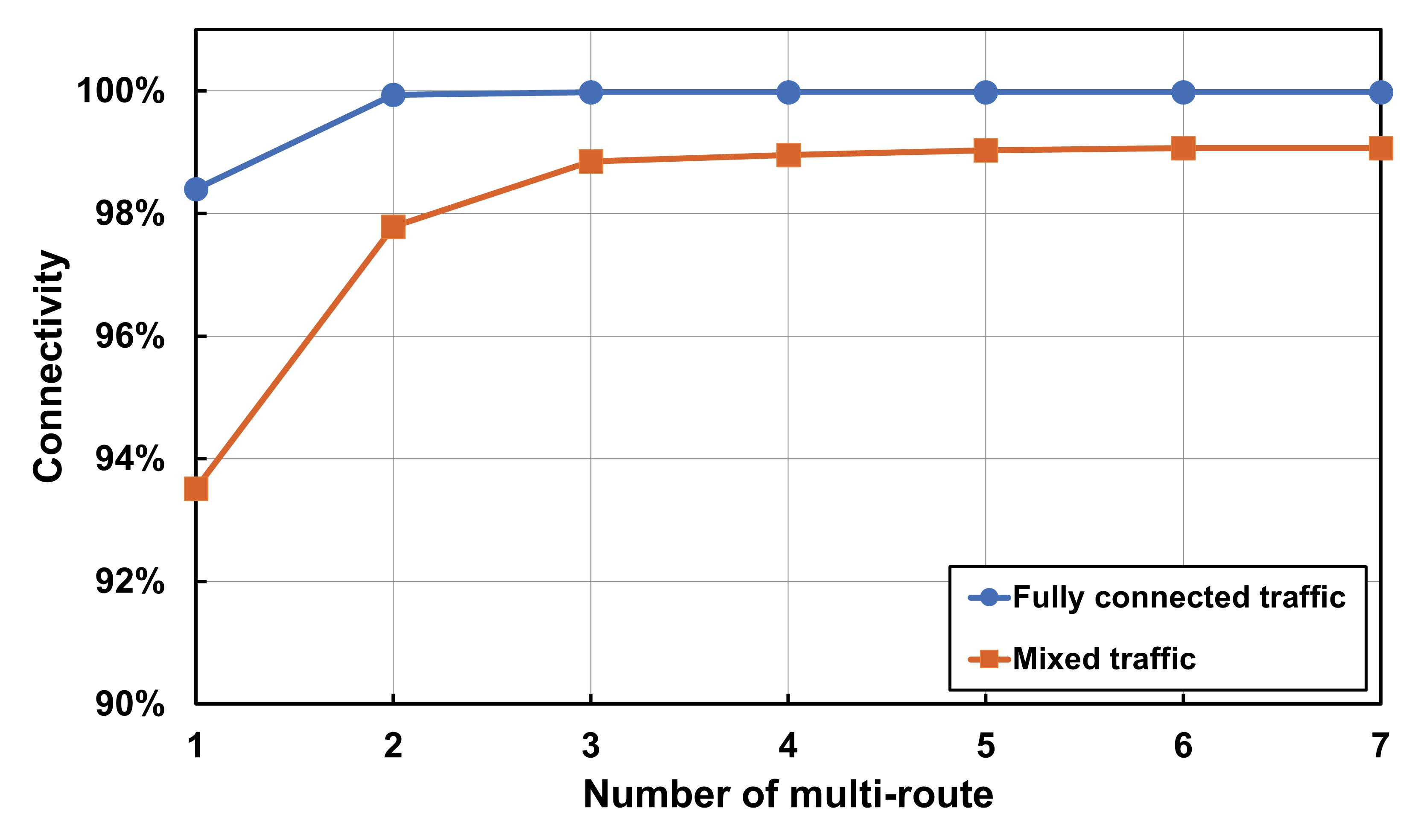}
    \caption{Effect of number of multi-routes on system connectivity.}
    \label{fig:n-multiroute}
\end{figure}

\begin{figure}[t]
    \centering
    \begin{subfigure}{0.48\textwidth}
        \includegraphics[width=\linewidth]{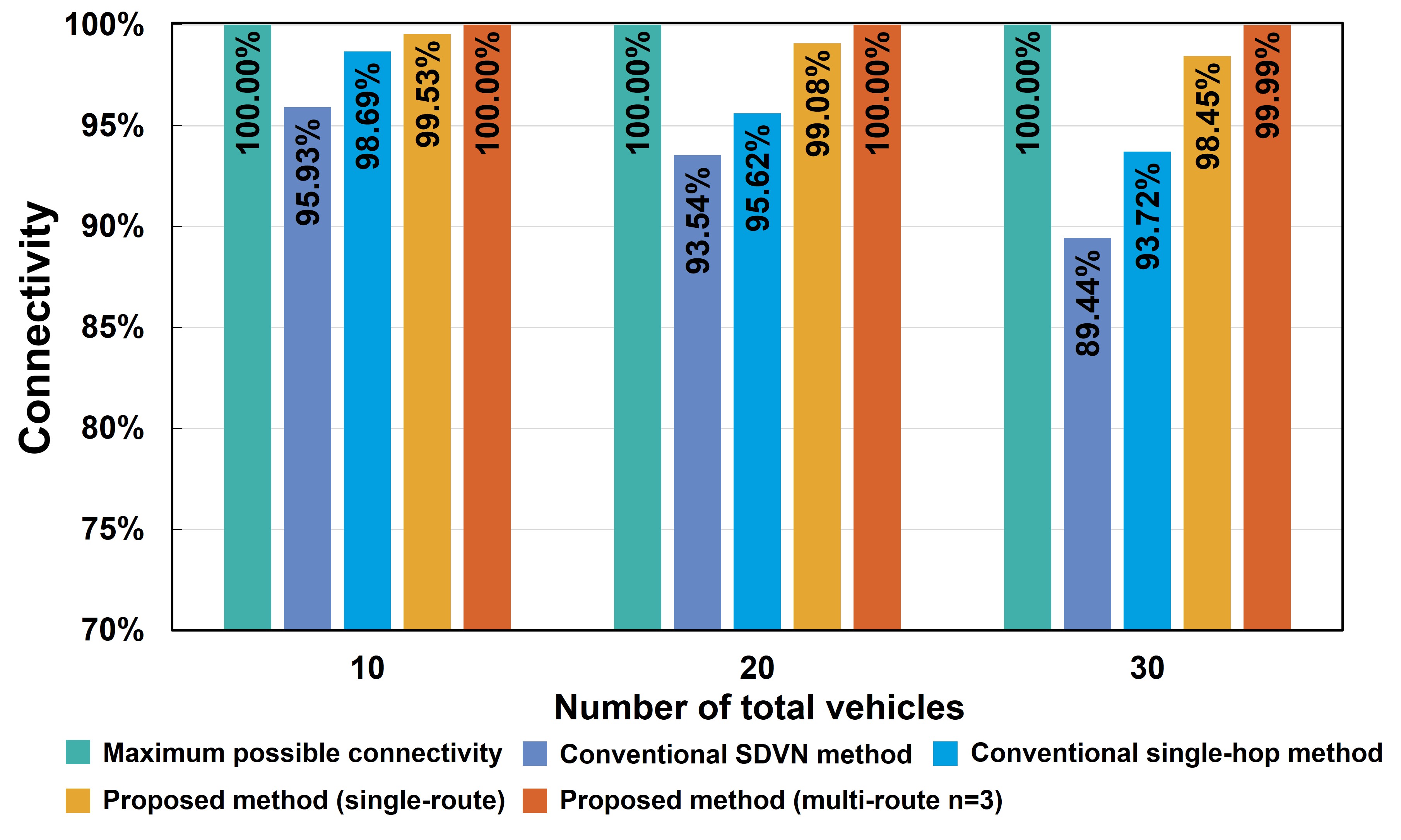}
        \caption{Fully connected traffic scenario}
        \label{fig:connected-connectivity}
    \end{subfigure}
    \hfill
    \begin{subfigure}{0.48\textwidth}
        \includegraphics[width=\linewidth]{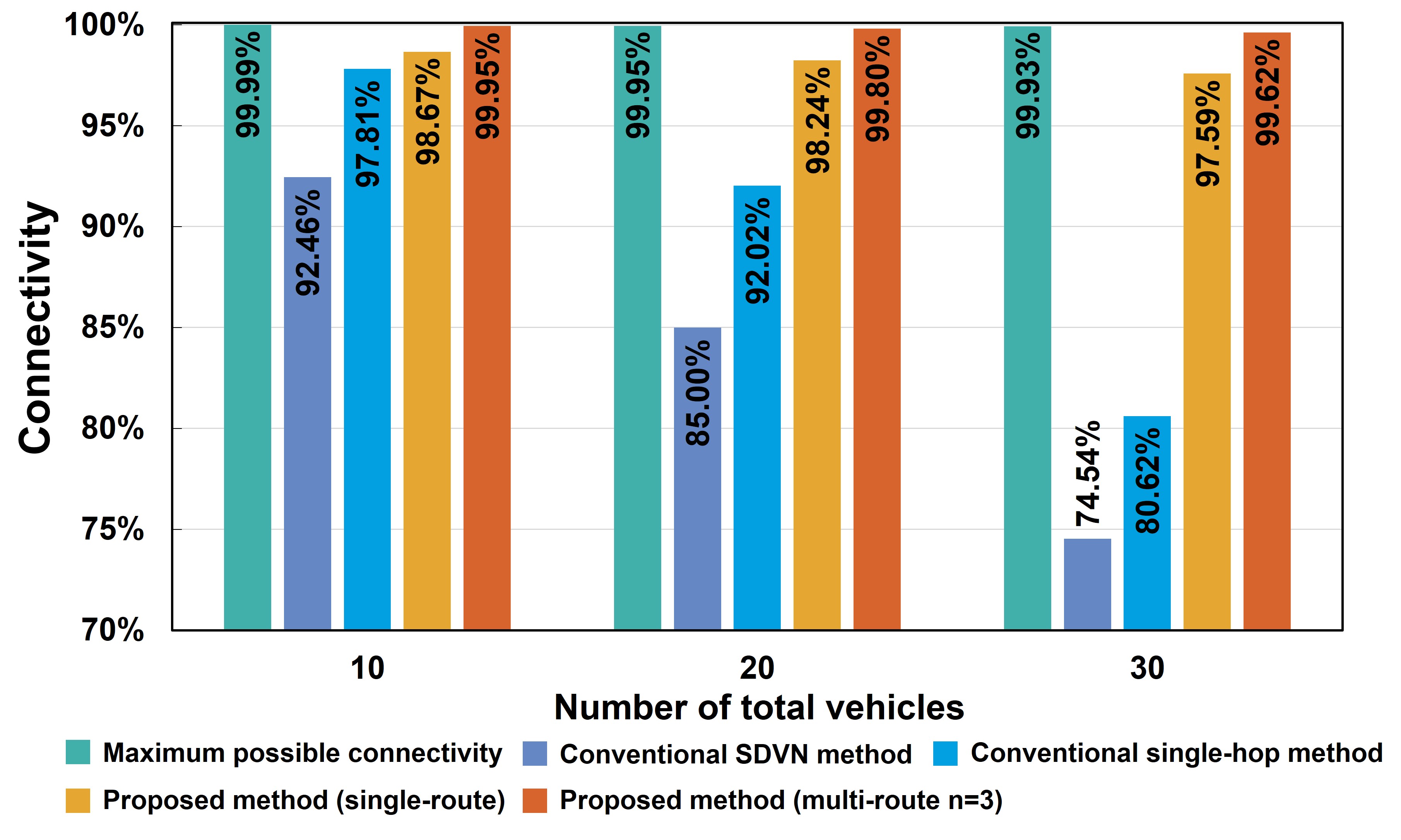}
        \caption{Mixed traffic scenario}
        \label{fig:mix-connectivity}
    \end{subfigure}
    \caption{Connectivity comparison between different routing methods.}
    \label{fig:connectivity-comparison}
\end{figure}

Leveraging the evaluation framework detailed in Section \ref{sec:sim-implement}, we conducted a comprehensive system evaluation of the proposed digital twin-enabled blockage-aware dynamic mmWave multi-Hop V2X communication system across all scenarios outlined in Section \ref{sec:scenario}. Our evaluation encompassed several key aspects of system performance. We examined the relationships between communication link weight hyperparameter on system connectivity, as shown in Fig. \ref{fig:connectivity-lambda}. We also analyzed the correlation between the trajectory prediction model's training loss and system connectivity, illustrated in Fig. \ref{fig:model-loss}, and investigated the impact of the number of multi-routes on overall connectivity, presented in Fig. \ref{fig:n-multiroute}. The simulation parameters, carefully selected for optimal performance, are presented in Table \ref{tab:sim-param}. To contextualize our findings, we conducted a comparative analysis of connectivity and throughput between our proposed method and conventional approaches. The results of this comparison are visualized in Fig. \ref{fig:connectivity-comparison}. These multifaceted assessments provide a thorough understanding of the system's performance, allowing us to evaluate the effectiveness of the proposed method in various operational contexts.

Fig. \ref{fig:connectivity-lambda} illustrates the impact of the communication link weight hyperparameter ($\lambda$) on system connectivity across different traffic scenarios. In fully connected traffic, connectivity remains consistently high (98\% to 98.5\%) regardless of $\lambda$ values. However, mixed traffic exhibits greater sensitivity to this parameter, with connectivity increasing from 95.7\% to 97.6\% as $\lambda$ increases from 0 to 1. A particularly notable jump in connectivity occurs when $\lambda$ increases from 0 to 0.2, marking the transition from a system that ignores blocking losses to one that considers them. These findings indicate that higher $\lambda$ values make the system place greater emphasis on mitigating blocking risks compared to LOS path loss, leading to better connectivity performance.

Fig. \ref{fig:model-loss} illustrates the relationship between the trajectory prediction model's training loss, measured by normalized mean square error (NMSE), and the resulting system connectivity. For both fully connected traffic and mixed traffic scenarios, we observe a clear inverse relationship between NMSE and connectivity. As the NMSE increases from 0.001 to 1, indicating higher prediction errors, the system connectivity decreases. In the fully connected traffic scenario, connectivity remains relatively high, starting at nearly 100\% for low NMSE and gradually declining to about 84\% as NMSE increased. In contrast, the mixed traffic scenario shows a more dramatic decrease in connectivity, starting at about 97\% for low NMSE and sharply dropping to approximately 68\% as NMSE increased. This steeper decline in the mixed traffic scenario suggests that accurate trajectory prediction is particularly crucial in more complex and diverse traffic environments, especially due to the increased likelihood of blockage. The results emphasize the dependency of model accuracy on system performance, as minimizing prediction errors in the trajectory model is essential for maintaining high levels of connectivity. Since the communication topology is determined based on trajectory predictions, reducing errors becomes especially important in mixed traffic conditions, where accurate predictions help preemptively mitigate potential blockages and ensure reliable communication.

Fig. \ref{fig:n-multiroute} illustrates the relationship between the number of multi-route paths and system connectivity for both fully connected and mixed traffic scenarios. In the fully connected scenario, connectivity starts at 98.40\% with one multi-route path, increases to 99.93\% with two routes, and stabilizes at 99.98\% from three routes onwards. In contrast, the mixed traffic scenario begins with lower connectivity at 93.52\% with one multi-route path and shows a gradual increase with the addition of routes, reaching 97.79\% with two routes and 98.85\% with three routes. However, after three routes, the increase in connectivity becomes minimal, stabilizing at around 99.06\% with six and seven routes. This suggests that while the mixed traffic scenario does benefit from more multi-route paths, the improvements in connectivity are less significant beyond three routes. Overall, the results highlight the critical role of repetition-based multi-route paths in enhancing connectivity, particularly in complex mixed traffic environments where maintaining reliable communication is essential.

\begin{table}[t]
    \centering
    \caption{Simulation parameters used in the system evaluation experiments.}
    \begin{tabularx}{1.0\linewidth}{|X|>{\centering\arraybackslash}p{1.2cm}|>{\centering\arraybackslash}p{1cm}|}
        \hline
        \textbf{Simulation Parameters} &  \textbf{Symbol} & \textbf{Value}\\
        \hline
        Millimeter wave center frequency & $f_{\text{mmWave}}$ & 60 GHz \\
        \hline
        Communication link weight hyperparameter & $\lambda$ & 1 \\
        \hline
        Trajectory prediction model training NMSE loss & $L_{NMSE}$ & 0.0046 \\
        \hline
        Number of multi-route & $n$ & 3 \\
        \hline
    \end{tabularx}
    \label{tab:sim-param}
\end{table}

The simulation parameters outlined in Table \ref{tab:sim-param} provide critical specifications for evaluating the proposed system. The millimeter wave center frequency is set to 60 GHz, following the WiGig standard \cite{cite:wigig} for demonstration purposes. The communication link weight hyperparameter ($\lambda$) is set to 1 to fully consider the effect of blockage risk in routing decisions. The trajectory prediction model's training NMSE loss value is recorded at 0.0046, indicating minimal prediction errors that support robust connectivity. Furthermore, the number of multi-routes is fixed at $n=3$, providing redundant paths for enhanced connectivity while maintaining reasonable computational complexity.

The connectivity comparison depicted in Fig. \ref{fig:connectivity-comparison} highlights the performance of various routing methods under two distinct traffic scenarios: fully connected and mixed traffic. 

In the fully connected traffic scenario (Fig. \ref{fig:connected-connectivity}), the maximum possible connectivity is fully achieved at 100\% across different network configurations. The proposed multi-route method excels by maintaining this near-perfect connectivity (99.99\% - 100\%) across all tested vehicle counts (10, 20, and 30 vehicles). Similarly, the proposed single-routing method demonstrates strong performance with 98.45\% connectivity at 30 vehicles. The conventional single-hop method initially performs well at 98.69\% but gradually declines to 93.72\% as the vehicle count increases to 30. The conventional SDVN method shows the poorest performance, with connectivity deteriorating from 95.93\% to 89.44\% as the network expands, highlighting its limitations in maintaining robust connectivity. 

For the mixed traffic scenario (Fig. \ref{fig:mix-connectivity}), the maximum possible connectivity is slightly lower, ranging from 99.93\% to 99.99\%. Again, the proposed multi-route method (n=3) demonstrates superior performance, achieving connectivity of 99.95\% at 10, 99.80\% at 20 vehicles, and a high 99.62\% at 30 vehicles. The proposed single-routing method remains robust with 97.59\% connectivity at 30 vehicles. However, the conventional single-hop method exhibits a steeper performance decline at higher vehicle densities, dropping to 82.05\%. The conventional SDVN method shows the most significant degradation, falling to 76.03\% with 30 vehicles.

To sum up, the results evaluation demonstrates that the proposed DT-enabled V2X multi-hop routing methods significantly outperform the conventional routing algorithm. The multi-route method maintains exceptional connectivity (99.62\% - 100\%) even under challenging conditions with high vehicle density and obstacles. While the single-route method also delivers reliable performance, conventional single-hop and SDVN methods struggle significantly, with connectivity falling to 80.62\% and 74.54\% respectively. These results indicate that the proposed methods, particularly the multi-route approach, can substantially improve the reliability and connectivity of real-world V2X networks.

\section{Conclusion}
\label{sec:conclusion}
In conclusion, this paper has demonstrated the effectiveness of DT technology in enhancing mmWave V2X communication reliability through a novel blockage-aware dynamic multi-hop routing system. By integrating mobility DTs with the multi-hop routing control plane, our proposed architecture enables informed routing decisions based on rich contextual data about vehicles, infrastructure, and potential signal blockages. Our DT-based blockage-aware routing algorithm, which combines high-precision DT data with trajectory prediction, achieves superior connectivity compared to conventional methods.

The proof-of-concept evaluation, conducted using a mobility DT of the Nishishinjuku area, demonstrated our system's exceptional performance. The proposed routing algorithm with multi-route technique maintained near-perfect connectivity (99.62\% to 100\%) even in challenging scenarios with high vehicle density and numerous obstacles. The single-route method also showed strong performance, further validating our approach. These results demonstrate that our DT-enabled system can significantly enhance the reliability and efficiency of mmWave V2X communications in environments with dynamic signal blockage.

The proposed system offers a promising V2X network control solution for future CAVs and ITS systems, particularly in scenarios requiring high data rates, low latency, and high reliability. Future work could explore developing more advanced DT-enabled routing algorithms to further enhance system performance or investigate system scalability. Additionally, more scenarios, such as NLOS communication in broader and more complex traffic conditions, could be considered. An outdoor proof-of-concept evaluation could also be conducted to validate performance in real-world environments.

\section*{Acknowledgment}
This work was partly supported by DENSO Corporation.

\bibliographystyle{bib/IEEEtran}
\bibliography{bib/bibliography}

\begin{thebibliography}{10}
\providecommand{\url}[1]{#1}
\csname url@rmstyle\endcsname
\providecommand{\newblock}{\relax}
\providecommand{\bibinfo}[2]{#2}
\providecommand\BIBentrySTDinterwordspacing{\spaceskip=0pt\relax}
\providecommand\BIBentryALTinterwordstretchfactor{4}
\providecommand\BIBentryALTinterwordspacing{\spaceskip=\fontdimen2\font plus
\BIBentryALTinterwordstretchfactor\fontdimen3\font minus \fontdimen4\font\relax}
\providecommand\BIBforeignlanguage[2]{{%
\expandafter\ifx\csname l@#1\endcsname\relax
\typeout{** WARNING: IEEEtran.bst: No hyphenation pattern has been}%
\typeout{** loaded for the language `#1'. Using the pattern for}%
\typeout{** the default language instead.}%
\else
\language=\csname l@#1\endcsname
\fi
#2}}

\bibitem{cite:v2x-services}
S.~Chen, J.~Hu, Y.~Shi, Y.~Peng, J.~Fang, R.~Zhao, and L.~Zhao, ``{Vehicle-to-Everything (v2x) Services Supported by LTE-Based Systems and 5G},'' \emph{IEEE Communications Standards Magazine}, vol.~1, no.~2, pp. 70--76, 2017.

\bibitem{cite:saej3016}
{SAE J3016}, ``{Taxonomy and Definitions for Terms Related to OnRoad Motor Vehicle Automated Driving Systems},'' Tech. Rep., Jun. 2018.

\bibitem{cite:mmWave-V2X}
K.~Sakaguchi, R.~Fukatsu, T.~Yu, E.~Fukuda, K.~Mahler, R.~Heath, T.~Fujii, K.~Takahashi, A.~Khoryaev, S.~Nagata, and T.~Shimizu, ``Towards mmwave v2x in 5g and beyond to support automated driving,'' \emph{IEICE Transactions on Communications}, vol. E104.B, 11 2020.

\bibitem{cite:cooperative-perception}
\BIBentryALTinterwordspacing
R.~Fukatsu and K.~Sakaguchi, ``Automated driving with cooperative perception using millimeter-wave v2v communications for safe overtaking,'' \emph{Sensors}, vol.~21, no.~8, 2021. [Online]. Available: \url{https://www.mdpi.com/1424-8220/21/8/2659}
\BIBentrySTDinterwordspacing

\bibitem{cite:mmwave-v2x-challenge}
L.~Liang, H.~Peng, G.~Y. Li, and X.~Shen, ``Vehicular communications: A physical layer perspective,'' \emph{IEEE Transactions on Vehicular Technology}, vol.~66, no.~12, pp. 10\,647--10\,659, 2017.

\bibitem{cite:3gpp.22.185}
3GPP, ``{Service requirements for V2X services},'' Technical Specification (TS) 22.185, Jul. 2020, {Version 16.0.0}.

\bibitem{cite:3gpp.22.186}
{3GPP}, ``{Enhancement of 3GPP support for V2X scenarios},'' Technical Specification (TS) 22.186, Jun. 2019, {Version 16.2.0}.

\bibitem{cite:v2x-broadcasting}
G.~Liu, Z.~Wang, J.~Hu, Z.~Ding, and P.~Fan, ``Cooperative noma broadcasting/multicasting for low-latency and high-reliability 5g cellular v2x communications,'' \emph{IEEE Internet of Things Journal}, vol.~6, no.~5, pp. 7828--7838, 2019.

\bibitem{cite:DT-trend}
\BIBentryALTinterwordspacing
M.~Singh, E.~Fuenmayor, E.~P. Hinchy, Y.~Qiao, N.~Murray, and D.~Devine, ``Digital twin: Origin to future,'' \emph{Applied System Innovation}, vol.~4, no.~2, 2021. [Online]. Available: \url{https://www.mdpi.com/2571-5577/4/2/36}
\BIBentrySTDinterwordspacing

\bibitem{cite:mobility-dt}
K.~Wang, Z.~Li, K.~Nonomura, T.~Yu, K.~Sakaguchi, O.~Hashash, and W.~Saad, ``Smart mobility digital twin based automated vehicle navigation system: A proof of concept,'' \emph{IEEE Transactions on Intelligent Vehicles}, pp. 1--14, 2024.

\bibitem{cite:rw-a-1}
E.~Zola, A.~J. Kassler, and W.~Kim, ``Joint user association and energy aware routing for green small cell mmwave backhaul networks,'' in \emph{2017 IEEE Wireless Communications and Networking Conference (WCNC)}, 2017, pp. 1--6.

\bibitem{cite:rw-a-2}
D.~Triantafyllopoulou, K.~Kollias, and K.~Moessner, ``Selfish routing and link scheduling in mmwave backhaul networks,'' in \emph{ICC 2023 - IEEE International Conference on Communications}, 2023, pp. 4200--4205.

\bibitem{cite:rw-a-3}
T.-Q. Bai, C.-Y. Huang, and Y.-K. Lee, ``Reliably route iot packets in software defined mmwave mesh networks,'' \emph{IEEE Networking Letters}, vol.~5, no.~1, pp. 50--54, 2023.

\bibitem{cite:rw-a-4}
C.~Samarathunga, M.~Abouelseoud, K.~Sakoda, and M.~Hashemi, ``Multi-hop routing with proactive route refinement for 60 ghz millimeter-wave networks,'' in \emph{2021 IEEE 93rd Vehicular Technology Conference (VTC2021-Spring)}, 2021, pp. 1--5.

\bibitem{cite:rw-a-5}
I.~Rasheed, F.~Hu, Y.-K. Hong, and B.~Balasubramanian, ``Intelligent vehicle network routing with adaptive 3d beam alignment for mmwave 5g-based v2x communications,'' \emph{IEEE Transactions on Intelligent Transportation Systems}, vol.~22, no.~5, pp. 2706--2718, 2021.

\bibitem{cite:rw-a-6}
Y.~Li, X.~Zhang, L.~Yan, and D.~K. Sung, ``An opportunistic routing protocol based on position information for beam alignment in millimeter wave vehicular communications,'' in \emph{ICC 2022 - IEEE International Conference on Communications}, 2022, pp. 267--272.

\bibitem{cite:rw-b-1}
D.~Kreutz, F.~M.~V. Ramos, P.~E. Veríssimo, C.~E. Rothenberg, S.~Azodolmolky, and S.~Uhlig, ``Software-defined networking: A comprehensive survey,'' \emph{Proceedings of the IEEE}, vol. 103, no.~1, pp. 14--76, 2015.

\bibitem{cite:rw-b-2}
Z.~He, J.~Cao, and X.~Liu, ``Sdvn: enabling rapid network innovation for heterogeneous vehicular communication,'' \emph{IEEE Network}, vol.~30, no.~4, pp. 10--15, 2016.

\bibitem{cite:rw-b-3}
R.~Dos Reis~Fontes, C.~Campolo, C.~Esteve~Rothenberg, and A.~Molinaro, ``From theory to experimental evaluation: Resource management in software-defined vehicular networks,'' \emph{IEEE Access}, vol.~5, pp. 3069--3076, 2017.

\bibitem{cite:rw-b-4}
X.~Ge, Z.~Li, and S.~Li, ``5g software defined vehicular networks,'' \emph{IEEE Communications Magazine}, vol.~55, no.~7, pp. 87--93, 2017.

\bibitem{cite:rw-b-5}
G.~Secinti, B.~Canberk, T.~Q. Duong, and L.~Shu, ``Software defined architecture for vanet: A testbed implementation with wireless access management,'' \emph{IEEE Communications Magazine}, vol.~55, no.~7, pp. 135--141, 2017.

\bibitem{cite:rw-b-6}
O.~Sadio, I.~Ngom, and C.~Lishou, ``Design and prototyping of a software defined vehicular networking,'' \emph{IEEE Transactions on Vehicular Technology}, vol.~69, no.~1, pp. 842--850, 2020.

\bibitem{cite:rw-b-7}
Z.~Li, T.~Yu, R.~Fukatsu, G.~K. Tran, and K.~Sakaguchi, ``Towards safe automated driving: Design of software-defined dynamic mmwave v2x networks and poc implementation,'' \emph{IEEE Open Journal of Vehicular Technology}, vol.~2, pp. 78--93, 2021.

\bibitem{cite:rw-b-8}
Z.~Li, K.~Wang, T.~Yu, and K.~Sakaguchi, ``{Het-SDVN: SDN-Based Radio Resource Management of Heterogeneous V2X for Cooperative Perception},'' \emph{IEEE Access}, vol.~11, pp. 76\,255--76\,268, 2023.

\bibitem{cite:coop-perception-for-dt}
B.~Lu, X.~Huang, Y.~Wu, L.~Qian, D.~Niyato, and C.~Xu, ``Cooperative perception aided digital twin model update and migration in mixed vehicular networks,'' \emph{IEEE Transactions on Intelligent Transportation Systems}, vol.~26, no.~2, pp. 2293--2308, 2025.

\bibitem{cite:3gpp.37.885}
3GPP, ``{Study on evaluation methodology of new Vehicle-to-Everything (V2X) use cases for LTE and NR},'' Technical Report (TR) 37.885, Jun. 2019, {Version 15.3.0}.

\bibitem{cite:dijkstra}
E.~W. Dijkstra, ``A note on two problems in connexion with graphs,'' \emph{Numerische mathematik}, vol.~1, no.~1, pp. 269--271, 1959.

\bibitem{cite:multiroute}
M.~Di~Renzo, ``On the achievable diversity of repetition-based and relay selection network-coded cooperation,'' \emph{IEEE Transactions on Communications}, vol.~62, no.~7, pp. 2296--2313, 2014.

\bibitem{cite:yen-algo}
J.~Y. Yen, ``Finding the k shortest loopless paths in a network,'' \emph{Management Science}, vol.~17, no.~11, pp. 712--716, 1971.

\bibitem{cite:rsu-height}
\BIBentryALTinterwordspacing
S.~Wang, J.~Huang, and X.~Zhang, ``Demystifying millimeter-wave v2x: towards robust and efficient directional connectivity under high mobility,'' in \emph{Proceedings of the 26th Annual International Conference on Mobile Computing and Networking}, ser. MobiCom '20.\hskip 1em plus 0.5em minus 0.4em\relax New York, NY, USA: Association for Computing Machinery, 2020. [Online]. Available: \url{https://doi.org/10.1145/3372224.3419208}
\BIBentrySTDinterwordspacing

\bibitem{cite:awsim}
{TIER IV}, ``{AWSIM: Open source simulator for self-driving vehicles},'' \url{https://github.com/tier4/AWSIM}.

\bibitem{cite:lstm}
S.~Hochreiter and J.~Schmidhuber, ``Long short-term memory,'' \emph{Neural computation}, vol.~9, no.~8, pp. 1735--1780, 1997.

\bibitem{cite:networkx}
A.~A. Hagberg, D.~A. Schult, and P.~J. Swart, ``Exploring network structure, dynamics, and function using {NetworkX},'' in \emph{Proceedings of the 7th Python in Science Conference (SciPy2008)}, G.~Varoquaux, T.~Vaught, and J.~Millman, Eds., Pasadena, CA USA, August 2008, pp. 11--15.

\bibitem{cite:ros2}
\BIBentryALTinterwordspacing
S.~Macenski, T.~Foote, B.~Gerkey, C.~Lalancette, and W.~Woodall, ``Robot operating system 2: Design, architecture, and uses in the wild,'' \emph{Science Robotics}, vol.~7, no.~66, p. eabm6074, 2022. [Online]. Available: \url{https://www.science.org/doi/abs/10.1126/scirobotics.abm6074}
\BIBentrySTDinterwordspacing

\bibitem{cite:sdvn-sota}
O.~Sadio, I.~Ngom, and C.~Lishou, ``Design and prototyping of a software defined vehicular networking,'' \emph{IEEE Transactions on Vehicular Technology}, vol.~69, no.~1, pp. 842--850, 2020.

\bibitem{cite:sdvn-geo-pos}
\BIBentryALTinterwordspacing
D.~K.~N. Venkatramana, S.~B. Srikantaiah, and J.~Moodabidri, ``Scgrp: Sdn‐enabled connectivity‐aware geographical routing protocol of vanets for urban environment,'' \emph{IET Networks}, vol.~6, no.~5, p. 102–111, Sept. 2017. [Online]. Available: \url{https://doi.org/10.1049/iet-net.2016.0117}
\BIBentrySTDinterwordspacing

\bibitem{cite:single-hop-sota}
Z.~Li, L.~Xiang, X.~Ge, G.~Mao, and H.-C. Chao, ``Latency and reliability of mmwave multi-hop v2v communications under relay selections,'' \emph{IEEE Transactions on Vehicular Technology}, vol.~69, no.~9, pp. 9807--9821, 2020.

\bibitem{cite:wigig}
T.~Nitsche, C.~Cordeiro, A.~B. Flores, E.~W. Knightly, E.~Perahia, and J.~C. Widmer, ``Ieee 802.11ad: directional 60 ghz communication for multi-gigabit-per-second wi-fi [invited paper],'' \emph{IEEE Communications Magazine}, vol.~52, no.~12, pp. 132--141, 2014.

\end{thebibliography}

\end{document}